\documentclass{article}
\usepackage{xcolor}

\definecolor{OliveGreen}{rgb}{0.5, 0.5, 0}
\def\R{{{\rm I}\!{\rm R}}}

\def\bsG{\mathbb K}
\def\sG{\mathbb K}
\def\cT{\mathcal T}
\def\eps{\epsilon}

\def\d{\,{\rm d}\,}

\def\R{{{\rm I}\!{\rm R}}}

\def\d{\,{\rm d}\,}

\input colordvi.sty
\usepackage{mathtools,color}
\usepackage{amssymb}
\usepackage{amsfonts}
\usepackage[us,12hr]{datetime}
\usepackage{fancyhdr}
\usepackage[active]{srcltx}
\usepackage{pdfsync} 
\DeclareSymbolFontAlphabet{\mathbb}{AMSb}
\usepackage{latexsym}
\usepackage{amsmath}
\usepackage{amsfonts}
\usepackage{mathbbol}
\usepackage{srcltx}
\usepackage {amsbsy}
\usepackage{amssymb}
\usepackage{epsfig}
\usepackage{latexsym}
\usepackage{amsmath}
\usepackage{amsfonts}
\usepackage{amsmath,amssymb,amsthm}

\newcounter{mycounter}[section]

\begin{document}

\title{\bf Heat conduction with ``aging"  memory}
\author{{\Large Sandra Carillo}   \\ Dipartimento di Scienze di Base e Applicate
    per l'Ingegneria \\        \textsc{Sapienza} Universit\`a di Roma,  Rome, Italy
                         \\ \& \\
                         ~I.N.F.N. - Sezione Roma1, Gr. IV\\ Mathematical Methods in NonLinear Physics,  Rome, Italy \\
        sandra.carillo\symbol{64}uniroma1.it\\
           \and
{\Large Claudio Giorgi} \\ Dipartimento di Ingegneria Civile, Architettura, Territorio, Ambiente e di Matematica \\ Universit\`a degli Studi di  Brescia,  Brescia, Italy
                         \\
        claudio.giorgi\symbol{64}unibs.it}      
 \date{}
\maketitle

\begin{abstract}
The term material with memory is generally used to indicate materials whose mechanical and/or thermodynamical behaviour depends not only on the process at the present time but also on the history of the process itself. 
Crucial in heat conductors with memory is the heat relaxation function which models the thermal  response of the material. The present study is concerned about a thermodynamical problem with memory ``aging"; that is, we analyze the temperature evolution
within a rigid heat conductor with memory whose relaxation function takes into account the aging of the material. In particular, we account for variations of the relaxation function due to a possible deterioration of the thermal response of the material related to its age.
\end{abstract}

\section{Introduction}
\setcounter{equation}{0}
Mechanical and/or thermodynamical behaviour of a material with memory depends not only on the process at the present time but also on the history of the process itself. 
 In particular, rigid thermodynamics with memory refers to the study of materials that exhibit rigid behavior (resistant to deformation) while also demonstrating memory effects in their thermodynamic properties. These materials respond not just to their current state, but also to their past history of stress, temperature, or deformation. Examples include shape-memory alloys, certain polymers, and glasses, which require advanced thermodynamic models to describe their non-equilibrium, path-dependent behaviors. Understanding these materials is crucial for developing applications in actuators, energy storage, and smart materials. The interest in such material is testified by a wide literature focussing on the mathematical properties of  the model \cite{new5-Hristov, new4-Hristov, GMZ}. Or concerned about how to model particular problem as, for instance, in \cite{new2-Deseri-et-al}. 
However, we may stress that the present investigation can be regarded as a needed to describe and investigate more complicated systems in which the thermal effects are combined with the other ones such as when the body is not assumed to be rigid, but also the formation is taken into account. Thermal effects with memory are important in many fields, such as mechanical engineering, materials science, and structural analysis, as temperature changes can induce stresses and strains in materials, which in turn can affect their mechanical behavior. A wide variety of results concerns models in which the combined effect of different physical phenomena need to be taken  into account. This is the case of so the called {\it smart materials}   or materials of biological origin. In \cite{new6-Hilton}, for instance, the two models of the thermo-elasticity and thermo- viscoelasticity are compared when the body is not considered to be rigid, as in our case, but an elastic or a viscoelastic response is  considered. Even more generally further phenomena can be incorporated in the model. In particular of great importance are magnetic media. A notable example of magnetic sensible media is represented by a gel whose magnetic sensibility is due to the injection of magnetic, sensible,  micro, or nanoparticles \cite{new7-cinesi2}.

Aging is also a time-dependent process; the macroscopic manifestations of aging, whatever the cause, may be modeled by a variation of the constitutive properties as time passes.
In materials with memory, the response of a non-aging substance to an external action changes with the passage of time independently of the moment in which the experiment is started. Instead, the properties of aging materials with memory change with the passage of time even in the absence of external agents. Hence two time scales are required to unambiguously describe the constitutive properties of an aging material. One time scale is needed to keep track of the time since the manufacturing of the material and the other time scale is needed to keep track of the time the material is under an external action.  Multiscaling models are very much up-to-date: they turn out to represent a useful tool to take into account different type of effects. Aim is to develop models capable two capture the memory effects at both the  microscopic as well as at the macroscopic level \cite{new9-Grmela}.

The problem here addressed to is within the framework of  rigid heat conduction 
with integral form memory. This theory originates from the works
of Coleman \cite{Coleman} and Gurtin and Pipkin \cite{GurPip}, based on the previous celebrated paper by  Cattaneo \cite{Cattaneo}.
Further investigations on the physical model concerning thermodynamics of 
materials with memory are due to   Coleman and Dill \cite{ColemanDill},   Giorgi and Gentili \cite{GioGen},  Amendola, Fabrizio and  Golden \cite{new4}.
Specifically, on the basis of the Gurtin-Pipkin model \cite{GurPip}, the thermodynamic
theory developed by  by Fabrizio, Gentili  and Reynolds \cite{FGR} is adopted. We provide some hints of this theory below.

Many other  non-integral form models  have been proposed for rigid heat conductors. For example, we recall the Green-Naghdi \cite{GN} and Quintanilla \cite{Quintanilla} models using the thermal displacement concept, the Tsou dual-phase lag model \cite{DYT,Quinta2002}, and the Burgers-type model \cite{GMZ,GZ} based on a rate-type constitutive equation.

\subsection{Aims of the paper}

The general strategy of our approach to the problem is introduced in Section 2. 
The aging effects are modeled by assuming that the time dependence of the heat flux relaxation function on $t$ (the current value) and $\tau$ (the past value) does not occur only through their difference $t-\tau$ (the so-called elapsed time), but involves $t$ and $\tau$ separately. In particular, we consider both the Cattaneo-Maxwell and Quintanilla models with aging memory and show that both lead to the same temperature evolution equation, where the difference is limited to the different  expression of the memory kernel.
The evolution equation associated with the Burgers-type model is proved to be slightly more complex than that obtained for the Maxwell and Quintanilla models. Indeed, an additional weak damping proportional to the rate of the relative temperature is present.

In Section 2 heat conduction models with aging memory are introduced and discussed, also providing some examples.
In Section 3 the temperature evolution of the models under investigation is connected to a 
second order integro-differential problem whose kernel involves the current and past time values separately. On use of the  free energy functional,  some a priori estimate
 are obtained. On the basis of the proved results, the existence of a unique strong solution is established in Section 4
borrowing the procedure devised in \cite{CDGP}.

\section{Heat conduction models with aging memory}
Let ${\bf x}\in\Omega\subset {\R}^3$ denote the current position within the 
heat conductor and $ t \in [0,+\infty)$  the time variable. Consider the  relative  temperature 
$$ u({\bf x},t):=\theta({\bf x},t) - \theta_0$$
 which is the  difference between the  absolute temperature $\theta$ and 
a fixed uniform reference temperature $\theta_0$,  for instance, the temperature of the surrounding environment  assumed not to be affected by the thermodynamical status of the conductor (see, e.g., \cite{ColemanDill}).
According to \cite{FGR} the thermodynamic state of the material at each point is
determined when the  temperature $u$ and its history $u^t$, 
  \begin{equation}\label{thetat}
 u^t ({\bf x},s): =   u ({\bf x}, t-s)~,
 \end{equation}
are given together with  the temperature  gradient $\bf g$ and its history ${\bf g}^t$,
\begin{equation}\label{g-bar-t0}
\displaystyle{{\bf g}({\bf x}, t):= \nabla u({\bf x}, t)~~~,~~{\bf g}^t({\bf x}, s): = {{\bf g}({\bf x},t-s) .}}
\end{equation}
or, alternatively, the integrated history of the temperature  gradient $\bar{\bf g}^t$ defined as
\begin{equation}\label{g-bar-t}
\displaystyle{\bar{\bf g}^t({\bf x},s): =\int_{t-s}^{t} {{\bf g}({\bf x},\tau) ~d \tau~.}}
\end{equation}
In \eqref{thetat}-\eqref{g-bar-t}  $t\in\R$ represents the current time, $\tau\in(-\infty,t]$ denotes an instant in the past, while $s=t-\tau\in[0,\infty)$ is usually referred to as {\it elapsed time}.

Hereafter we restrict our attention to linear models of heat conduction.  In particular, we assume that the heat  flux vector ${\bf q}$  is given by   
\begin{equation}\label{lin_q}
{\bf q} ({\bf x}, t) =-{\int^{t}_{-\infty} { k( t-\tau) {\bf g}({\bf x}, \tau) ~d \tau}}
=-{\int_{0}^{\infty} { k( s) {\bf g}^t({\bf x}, s) ~d s }}
\end{equation}
where the memory kernel ${k}(t)$ is referred to as {\it heat flux relaxation function} (see \cite{FGR}). This constitutive equation 
was first obtained in the linearized Gurtin-Pipkin theory \cite{GurPip}.
Moreover, \eqref{lin_q} is a generalization of the Maxwell-Cattaneo model,
\begin{equation}\label{MC_model}
\xi_0{\bf q}_t ({\bf x},  t) +{\bf q} ({\bf x},  t)=
- \kappa_0 {\bf g}({\bf x}, t) ,
\end{equation}
(the subscript $_t$ denotes partial differentiation with respect to time).
Unlike Fourier's law, this model takes relaxation into account and hence $\xi_0$ is referred to as {\it relaxation time}.
Letting  ${\bf q}({\bf x}, t_0) = {\bf q}_0({\bf x})$ and solving \eqref{MC_model} for $t>t_0$
in  the unknown ${\bf q}$ we find
\[{\bf q}({\bf x}, t) ={\bf q}_0({\bf x} ) \exp\big[-(t-t_0)/\xi_0\big] -\frac{\kappa_0}{\xi_0} \int_{t_0}^t \exp\big[-(t-\tau)/\xi_0\big] {\bf g}({\bf x}, \tau) \,d\tau, \]
 Letting $t_0 \to -\infty$ we obtain
\[ {\bf q}({\bf x}, t)  =-\frac{\kappa_0}{\xi_0} \int_{-\infty}^t \exp\big[-(t-\tau)/\xi_0\big] {\bf g}({\bf x}, \tau) \,d\tau,\]
and  the change of variable $t- \tau=s$ yields
\[  {\bf q}({\bf x}, t)  =-\frac{\kappa_0}{\xi_0}
\int_0^\infty \exp(-s/\xi_0)\,{\bf g}^t({\bf x}, s) \,ds, \]
which provides a special case of \eqref{lin_q} by setting $k(s)=\frac{\kappa_0}{\xi_0}\exp(-s/\xi_0)$.

Finally, letting $k_0:= k(0)$ take on a finite value, an integration by parts of \eqref{lin_q} leads to the alternative integral form
\[  {\bf q}({\bf x}, t)  =k_0{\bf g}({\bf x}, t)+ 
\int_0^\infty  k'(s)\,\bar{\bf g}^t({\bf x}, s) \,ds,\]
where the prime denotes differentiation with respect to the (unique)  argument. Otherwise, a derivation of \eqref{lin_q}  with respect to time $t$ gives
\begin{equation}{\bf q}_t ({\bf x}, t) =-k_0{\bf g}({\bf x}, t) -{\int^{t}_{-\infty} { k'( t-\tau) {\bf g}({\bf x}, \tau) ~d \tau}}=
-k_0{\bf g}({\bf x}, t) -{\int_{0}^{\infty} { k'( s) {\bf g}^t({\bf x}, s) ~d s}}
\label{q_t}\end{equation}

An alternative approach starts from the so-called Quintanilla constitutive model. 
Borrowing from  the Green-Naghdi theory, the Quintanilla model involves the temperature displacement $\alpha({\bf x}, t)$ such that $\alpha_t({\bf x}, t)=\theta({\bf x}, t)$, namely
\[
\xi_0{\bf q}_{t} ({\bf x}, t) +{\bf q} ({\bf x}, t)=-h_0\nabla \alpha({\bf x}, t)- \kappa_0 {\bf g}({\bf x}, t) .
\]
Taking advantage of our notation of the integrated history \eqref{g-bar-t}, we can identify the gradient of temperature displacement with the integrated history of $\bf g$ at time $t$,
\[\nabla \alpha({\bf x}, t)=\int_0^t{\bf g}({\bf x},\tau)\d\tau=\bar{\bf g}^t({\bf x}, t).\]
Accordingly we can write
\[\xi_0{\bf q}_{t} ({\bf x}, t) +{\bf q} ({\bf x}, t)=-h_0\bar{\bf g}^t({\bf x}, t)- \kappa_0 {\bf g}({\bf x}, t). \]
We can solve it in  the unknown ${\bf q}$ so obtaining the integral equation
\[ {\bf q}({\bf x},t)  =- \int_{-\infty}^t \frac{1}{\xi_0}\exp\big[-(t-\tau)/\xi_0\big] [h_0\bar{\bf g}^\tau({\bf x},\tau)+\kappa_0{\bf g}({\bf x},\tau)] \,d\tau,\]
from which it follows
\[ {\bf q}({\bf x},t) =-{h_0}\bar{\bf g}^t({\bf x},t) + \int_{-\infty}^t \frac{\kappa_0-\xi_0h_0}{\xi_0}\exp\big[-(t-\tau)/\xi_0\big] {\bf g}({\bf x},\tau)\,d\tau.\]

Taking advantage of its rate-type form 
(see \cite{GMZ}), namely
\[
\xi_0{\bf q}_{tt} ({\bf x},  t) +{\bf q}_t ({\bf x},  t)=-h_0{\bf g}({\bf x}, t)
- \kappa_0 {\bf g}_t({\bf x}, t) ,
\]
we can solve it in  the unknown ${\bf q}_t$ so obtaining the integral equation
\[ {\bf q}_t({\bf x}, t)  =- \int_{-\infty}^t \frac{1}{\xi_0}\exp\big[-(t-\tau)/\xi_0\big] [h_0{\bf g}({\bf x}, \tau)+\kappa_0{\bf g}_\tau({\bf x}, \tau)] \,d\tau,\]
from which it follows
\[ {\bf q}_t({\bf x}, t) =-\frac{\kappa_0}{\xi_0}{\bf g}({\bf x}, t) + \int_{-\infty}^t \frac{\kappa_0-\xi_0h_0}{\xi_0}\exp\big[-(t-\tau)/\xi_0\big] {\bf g}({\bf x}, \tau)\,d\tau.\]
Assuming $\kappa_0>\xi_0h_0$ (see \cite[Proposition 5.3]{GZ}) and identifying 
\[k_0=\frac{\kappa_0}{\xi_0}, \qquad k(s)=\frac{\kappa_0-\xi_0h_0}{\xi_0}\exp\big[-s/\xi_0\big]\]
we recover \eqref{q_t}, so proving that both Cattaneo-Maxwell and Quintanilla models lead to the same form of integral constitutive equation. Nevertheless, as to the integral equation obtained from the Quintanilla model, we remark that the value of $k_0$ does not match with $k(0)$, the initial value of the memory kernel.

According to \cite{GMZ}, a Burgers-type model for heat conductors is characterized by the rate-type equation
\begin{equation}
\xi_0{\bf q}_{tt}+\nu_0{\bf q}_t+{\bf q}=-h_0{\bf g}- \nu_0 \kappa_0{\bf g}_t,\qquad \xi_0,\nu_0>0.\label{Burg}
\end{equation}
Thermodynamic consistency with the CD inequality is guaranteed by (see \cite[Sect.~3.2.2]{GMZ})
\begin{equation}
\label{TD_consistency}
h_0>0,\qquad\nu^2_0\kappa_0-\xi_0h_0\ge0.
\end{equation}
Since $\xi_0,\nu_0>0$  we can choose $\mu_1,\mu_2>0$ such that
\[\begin{cases}
 \mu_1+\mu_2=\nu_0      \\
 \mu_1\mu_2=\xi_0     
\end{cases}\]
and equation \eqref{Burg} can be rewritten as
\[
\mu_1(\mu_2{\bf q}_{t}+{\bf q})_t+\mu_2{\bf q}_t+{\bf q}=-h_0{\bf g}- \nu_0 \kappa_0{\bf g}_t.
\]
By solving it in  the unknown $z=\mu_2{\bf q}_{t}+{\bf q}$ we obtain the integral equation
\[ z(t):= \mu_2{\bf q}_{t}(t)+{\bf q}(t)  =- \int_{-\infty}^t \frac{1}{\mu_1}\exp\big[-(t-\tau)/\mu_1\big] [h_0{\bf g}(\tau)+\nu_0\kappa_0{\bf g}_\tau(\tau)] \,d\tau,\]
from which it follows
\begin{equation}
\label{Burg_integral}
{\bf q}_{t}(t)+\frac1{\mu_2}{\bf q}(t)  =-\frac{\nu_0\kappa_0}{\xi_0}{\bf g}(t) + \int_{-\infty}^t \frac{\nu_0\kappa_0-\mu_1h_0}{\mu_1\xi_0}\exp\big[-(t-\tau)/\mu_1\big] {\bf g}(\tau)\,d\tau.
\end{equation} 
Within the integral term we are allowed to identify 
\begin{equation}k_0=\frac{\nu_0\kappa_0}{\xi_0}, \qquad k(s)=\frac{\nu_0\kappa_0-\mu_1h_0}{\xi_0}\exp\big[-s/\xi_0\big]\label{k_consistence}\end{equation}
provided that $\nu_0\kappa_0-\mu_1h_0>0$.
This condition can be written as
\[\nu^2_0\kappa_0-\xi_0h_0>\mu_1^2h_0\]
and turns out to be stronger than \eqref{TD_consistency}.

\subsection{Thermodynamics of heat conduction models with memory}

 The Clausius-Duhem inequality,  representing the second law of 
thermodynamics,  can be written as
\begin{equation}\label{m1}
\displaystyle{\eta}_t  \ge \frac 1 \theta { e}_t  + 
 {\bf q} \cdot \frac{\nabla\theta}{ \theta^2}
 \end{equation}
wherein, respectively, $\eta$ denotes the specific entropy and $e$ the specific internal energy. Note that $\nabla\theta=\nabla u={\bf g}$.
Within the linear theory of heat conductors with memory,  Fabrizio, Gentili  and Reynolds  \cite{FGR} introduce 
the free pseudoenergy
$$\zeta:=\theta_0 (e- \theta_0 \eta),$$
where $\theta_0$ is a given uniform reference temperature and
\[e= \hat e \left(\theta(t),\theta^t, \bar{\bf g}^t \right), \qquad \eta= \hat \eta \left(\theta(t),\theta^t, \bar{\bf g}^t \right),
\qquad
\zeta= \hat \zeta \left(\theta(t),\theta^t, \bar{\bf g}^t\right) ~.
\]
In  \cite{FGR} it is proved that the approximate canonical free pseudoenergy $\zeta(u(t), u^t,\bar{\bf g}^t )$, for small values of the relative temperature $u(t)$,
enjoys most of the properties that characterize a canonical free energy even if it is not dimensionally homogeneous to such a potential. In particular, the inequality
(\ref{m1}) reduces to
\begin{equation}\label{m2}
{{ \tilde\zeta}_t} \le {{ \tilde e}_t ~ u} +  {{\bf q} 
\cdot {\bf g}}~.
 \end{equation}
 where the approximate internal energy  $\tilde e$  is assumed  to be linear
    \begin{equation}\label{1b}
\tilde e ({\bf x}, t) =  \alpha_0 u ({\bf x}, t)~~.
 \end{equation}
For sake of simplicity, the specific heat $ \alpha_0$ is  assumed  to be positive and independent on the position within the heat conductor.

Following the arguments set out \cite{GioGen,FGR}, thermodynamic consistency is achieved by assuming that the heat flux relaxation function satisfies the following  requirements (often referred to as Graffi's conditions)
\begin{equation}\label{k}
{k}(t)>0,\qquad {k}'(t)\le 0, \qquad {k}''(t)\ge 0,\qquad
t\in(0,\infty).
\end{equation}
and
\begin{equation}\label{kt}
{k} \in L^1(0,T)\cap C^2(0,T) ~~ \forall T\in \R^+~.
\end{equation}
Note that inequalities (\ref{k}) together with $\alpha_0>0$,  
guarantee   that  
$$\tilde\zeta_G(u(t), u^t,\bar{\bf g}^t )=\alpha_0|u(t)|^2-\frac12\int_0^\infty k'(s)\,|\bar{\bf g}^t(s)|^2 \,ds$$
 represents a free pseudoenergy which satisfies  
Clausius-Duhem inequality  (\ref{m2}).
As it is well known, $\tilde\zeta_G$ is not unique (up to an additive constant). Indeed, there are many expressions of the free pseudoenergy $\tilde\zeta$ since there is  a convex set of functionals  which satisfies the inequality (\ref{m2}) (see, for instance, \cite{AMCA2004,AFG09}).

\subsection{Temperature evolution}
Let $r({\bf x},t)$ denote the energy supply. The energy balance for a rigid heat conductor  is given by
\begin{equation}\label{eq00}
e_t({\bf x},t)= -\nabla \cdot {\bf q}({\bf x}, t) + r({\bf x},t)~.
\end{equation}
Upon replacing the expressions (\ref{lin_q})  and (\ref{1b}) we obtain the equation which models the temperature evolution within
the conductor
\[
\displaystyle{ \alpha_0 u_t({\bf x},t)=  \int_{0}^{\infty} { k(s)~ {\Delta  u} \,({\bf x}, t-s) ~d s + r({\bf x},t)~ .}}
\]
More conveniently, it can be rewritten as  
\begin{equation}\label{eq3}
\displaystyle{ \alpha_0 u_t({\bf x},\tau) =  \int_{0}^{t} { k(t-\tau)~ {\Delta  u} \,({\bf
x}, \tau) ~d \tau\, + {f}({\bf x},t)~ ,}}
\end{equation}
where  the term 
$${f}({\bf x},t)=\int_{-\infty}^{0}  k(t-\tau)~ {\Delta  u} \,({\bf x}, \tau) ~d \tau+r({\bf x},t)$$
 denotes an external source term which also 
includes the history of the material up to $t=0$.

 If in addition to \eqref{kt} we assume the further regularity condition 
 \begin{equation}  k' \in L^1 (0,T),~ \forall T\in \R^+,\end{equation}
then the initial value of the heat-flux relaxation function, $k_0 := k(0)$, also called {\it initial heat flux relaxation coefficient},  turns out to be finite. This allows us to differentiate \eqref{eq3} and obtain
\begin{equation}\label{eq_main}
\displaystyle{ \alpha_0 u_{tt}({\bf x},t) = k_0~ {\Delta  u} \,({\bf x}, t)+ \int_{0}^{t} {  k'(s)~ {\Delta  u} \,({\bf x}, t-s) ~d s\, + {f}_t({\bf x},t)~ ,}}
\end{equation}
As can be easily verified, this evolution equation also corresponds to the constitutive model \eqref{q_t}. Therefore, both Cattaneo-Maxwell and Quintanilla models (in their integral form) lead to the same temperature evolution equation where the difference is limited to the different expression of the memory kernel $k$.

The evolution equation associated with the integral Burgers-type model \eqref{Burg} is slightly more complex than \eqref{eq_main}. Indeed, by adding \eqref{eq00} with its time derivative multiplied by $\mu_2$ we have
\[
e_t({\bf x},t)+\mu_2 e_{tt}({\bf x},t)= -\nabla \cdot [{\bf q}({\bf x}, t)+\mu_2{\bf q}_t({\bf x}, t)] + r({\bf x},t)+\mu_2 r_t({\bf x},t)~.
\]
Then, applying \eqref{Burg_integral}, \eqref{k_consistence} and  \eqref{1b}  we obtain
\[
\alpha_0u_{tt}({\bf x},t)+\frac{\alpha_0}{\mu_2}u_t({\bf x},t)  =-{k_0}\Delta u({\bf x},t) + \int_{0}^t k'(t-\tau)\Delta u({\bf x},\tau)\,d\tau +F({\bf x},t).
\]
where
\[F({\bf x},t)= \int_{-\infty}^{0}  k'(t-\tau)~ {\Delta  u} \,({\bf x}, \tau) ~d \tau+r_t({\bf x},t)+\frac{1}{\mu_2}r({\bf x},t).
\]

\subsection{Aging effects}

When { aging effects} are modeled, it can be assumed  
that  the dependence of the heat flux relaxation function $k$ on $t$ and $\tau$ does not occur only through their difference $t-\tau$ (as in \eqref{lin_q}) but involves $t$ and $\tau$ separately, namely 
${k}(t,\tau)$ (see, for instance, \cite[Ch.\,8]{MG_book} and \cite{CDGP}). 
Aging effects are taken into account by modifying \eqref{lin_q}  as follows (the dependence on $\bf x$ is understood and not written)
\[
\displaystyle{
{\bf q}\, (t) =  -\int^{t}_{-\infty}  {k}(t,\tau){\bf g}(\tau)d\tau}.
\]
and the classical (without aging) expression is recovered by simply assuming that 
\begin{equation*}
{k}(t,\tau)={k}(t-\tau), \qquad\tau\leq t.
\end{equation*}
 In particular, we obtain the following correspondence 
 \begin{equation*}{k}_\tau(t,\tau) =-{k}'(t-\tau), \qquad{k}(t,t)={k}_0,
 \end{equation*}
wherein the subscript $_\tau$ indicates partial derivative with respect to $\tau$. Hence, to account for memory aging  \eqref{q_t} is modified as follows
\begin{equation}\label{1.2bis}
\displaystyle{
{\bf q}_t (t) = -{k}(t,t){\bf g}(t) +\int^{t}_{-\infty}  {k}_\tau(t,\tau){\bf g}(\tau)d\tau
}.
\end{equation}

In view of further applications, it turn out to be convenient to introduce the scalar function  $\bsG$, defined on $\R\times\R^+$ as
$$\bsG(t,s)={k}(t, t-s),$$
and hence
\[
 \bsG_s(t,s)= {k}_s(t, t-s)=  -{k}_\tau(t, \tau), \qquad \tau=t-s.
\]
Accordingly \eqref{1.2bis} becomes
\begin{equation}\label{1.3}
\displaystyle{
{\bf q}_t(t) = -{\bsG}_0(t)
{\bf g}(t) -\int_{0}^{\infty}  {\bsG}_s(t,s){\bf g}^t(s)ds
}.
\end{equation}
where $\bsG_0 (t):=\bsG(t,0)={k}(t, t)$, and 
\begin{equation*}{\bsG}(t,s) =- {\bsG}_0(t) -
\int_{0}^{s} {  {\bsG}_\xi(t,\xi) ~d \xi  ,\qquad  \lim_{s\to +\infty}{\bsG}(t,s)=0.}
 \end{equation*}
Owing to \eqref{1.3}, the evolution equation \eqref{eq_main} is rewritten as
\begin{equation}\label{eq_aging}
\displaystyle{ \alpha_0 u_{tt}({\bf x},t) =\bsG_0 (t)~ {\Delta  u} \,({\bf x}, t)+ \int_{0}^{t} {  {\bsG}_s(t,s)~ {\Delta  u} \,({\bf x}, t-s) ~d s\, + {f}_t({\bf x},t)~ ,}}
\end{equation}

According to the physics of the model  (for viscoelastic materials, see assumptions M1-M4 in \cite{Carillo_Giorgi}),
 the relaxation kernel $\sG$ and its derivatives must satisfy
   \begin{equation}\label{G}
    \sG(t,s)>0,\quad {\sG}_s(t,s)\le 0, \quad {\sG}_{ss}(t,s)\ge 0,\quad
   ( t,s)\in\R\times\R^+,
    \end{equation}
      \begin{equation}\label{G_extra}
   {\sG}_{t}(t,s)+{\sG}_{s}(t,s)\le 0,\quad {\sG}_{ts}(t,s)+{\sG}_{ss}(t,s)\ge 0\quad ( t,s)\in\R\times\R^+.
    \end{equation}
Since ${\sG}(t,s)$ reduces to $k(s)$ when aging is neglected, assumptions \eqref{G} correspond to classical Graffi's conditions \eqref{k}, whereas  \eqref{G_extra} boils down to \eqref{G}.

\subsubsection*{Examples}
A typical example of aging memory kernel is given by
\[
 {\sG}(t,s)= \exp\Big(-\frac{s}{\eps(t)}\Big) ,
\]
where $\eps\in C^1(\R,\R^+)$ satisfies
$$
 \eps'(t)\leq 0, \quad\forall t\in\R.
$$
It is easy to verify that assumptions \eqref{G}, \eqref{G_extra}  are complied.

Another example is obtained by a suitable rescaling  of a (nonnegative) non-increasing
function $k$. Given $t_0>0$ and $\eps\in C^1(\R,\R^+)$ satisfying
$$
\eps'(t)\leq 0, \quad\forall t\in\R,
$$
we define
  \[
 \sG(t,s)=\frac{1}{\eps(t)}\,k\Big(\frac{s}{\eps(t)}\Big)  
   \]
In particular, for all $t\in\R$ 
we get
$$
\int_0^\infty  \sG(t,s)\d s=\int_0^\infty k(y)\d y=\mu<\infty.
$$
Accordingly, if $\eps(t)\to 0$ as $t\to \infty$, we obtain the distributional convergence
$\lim_{ t\to \infty}\sG(t,\cdot)= \mu \delta_0$,
where $\delta_0$ denotes the Dirac mass at $0^+$.
 For definiteness, we take
$k(s)={\rm e}^{-s}$ and $\eps(t)=\alpha/t$, $\alpha>0$,
in which case
  \[
  \sG(t,s)=  \frac{t}{\alpha}\exp\Big(-\frac{st}{\alpha}\Big)   
    \]
This relaxation kernel  complies with assumptions \eqref{G}, \eqref{G_extra} provided that $t\ge\sqrt{2\alpha}$.

 \rm
\section{The problem}
\setcounter{equation}{0}
The problem under investigation models the evolution of (relative) temperature in rigid heat conductors with time-dependent memory.  For sake of simplicity, the body  here considered is one-dimensional. Let $\alpha_0>0$, $\Omega =(0,1)$ and  ${\cal Q}:=\Omega \times (0,T)$, $T>0$. 
After applying \eqref{eq_aging} with $F=f_t$, we consider the following linear integro-differential equation in  ${\cal Q}$, 
\begin{equation}\label{eql2}
\displaystyle{ \alpha_0 u_{tt}({ x},t) = \sG(t,0) {u}_{xx} ({ x}, t)+ \int_{0}^{t} { \sG_{s}(t,s)~ { u}_{xx}({ x}, t-s) ~d s\, + {F}({ x},t)~ ,}}
\end{equation}
together with the initial and boundary conditions
\begin{equation} \label{ic+bc-G}
\begin{split}
&{u}(\cdot,0)=u_0,\quad {u}_t(\cdot,0)=u_1,\qquad {\text{\rm in }} \Omega \\
& {u}(0,\cdot)=  u(1,\cdot)=0,\ \quad\qquad\qquad  {\text{\rm in }}
 (0,T)\,.
 \end{split}
\end{equation}
In addition, the kernel, $\sG:{\cT}\to\R$, $\cT=[0,T]^2$, is supposed to satisfy  \eqref{G} and \eqref{G_extra}.

Borrowing the procedure devised in \cite{CDGP}, problem  (\ref{eql2})-(\ref{ic+bc-G}) admits a {unique strong solution}. In particular, the following result holds.

\bigskip\noindent
{\bf Lemma 3.1}. \it
Denote by ${u}$ the unique solution admitted to the problem \eqref{eql2}-\eqref{ic+bc-G} with 
 \begin{equation}\label{data}
u_0 \in H^1_0(\Omega), \quad u_1 \in L^2(\Omega), \quad  F\in L^2({\cal Q}).
\end{equation}
Let \eqref{G}-\eqref{G_extra} hold. Then for all $t\in[0,T]$ the following estimate is obtained
\begin{equation}
\begin{split}\displaystyle 
&{\frac12} \int_{\Omega} \sG(t,t)\, \vert{u}_x(t)\vert^2\, dx + {\frac12}
\int_{\Omega} \vert{u}_t(t)\vert^2\, dx\\
&\le {\frac12} \int_{\Omega} \sG(0,0)\, \vert{u}_{0x}\vert^2\, dx
+ {\frac12} \int_{\Omega} \vert{u}_1\vert^2\, dx 
+ \int_{\Omega} \int_0^t   F(\tau)\,{u}_\tau(\tau)\, dx \,d\tau .
\label{ineq-lemma1} 
\end{split}\end{equation}  
\rm

\begin{proof} First of all,  add and subtract to equation  (\ref{eql2}) the term 
  $$ \int_0^t {\sG}_s(t,s) {u}_{xx}(t) ds= \left[\sG(t,t)-\sG(t,0)\right]  {u}_{xx}(t)~.$$
The result can be written in the equivalent form
\begin{equation} \label{eql1}
   {u}_{tt} - {\sG}(t,t){u}_{xx} + \int_0^t {\sG}_s(t,s)\left[{u}_{xx}(t)-{u}_{xx}(t-s)\right] ds = F.
\end{equation}    
When equation (\ref{eql1}) is multiplied  by ${u}_t$,
after integration over $\Omega$   it follows 
\begin{equation}\label{form2}
\begin{array}{cl@{\hspace{0.5ex}}c@{\hspace
{1.0ex}}l} \displaystyle{ \frac12 {\frac d{dt}}\int_{\Omega}\vert{u}_t(t)\vert^2 dx
\,+\,\int_{\Omega}{\sG}(t,t) {u}_x(t)\, {u}_{xt}(t)\, dx \,+}
\\   \displaystyle{ \!\!\!\!\!\!\!\!\!\!\!\!\!\!\!\!
-\int_{\Omega} {u}_{xt}(t) \, \int_0^t {\sG}_s(t,s)
\,\left[{u}_{x}(t)-{u}_{x}(t-s)\right]ds \, dx = \int_{\Omega}
F {u}_t(t)\, dx.}
\end{array}\end{equation}

Since ${\frac d{dt}}{\sG}(t,t)= [{\sG}_{t}+{\sG}_{s}](t,s)\vert_{s=t}$, it follows 
\begin{equation}\label{form3}
\begin{array}{cl@{\hspace{0.5ex}}c@{\hspace
{1.0ex}}l}   
\displaystyle{{\frac12} {\frac d{dt}}\int_{\Omega}\vert{u}_t\vert^2 dx
\,+\,{\frac12} {\frac d{dt}} \int_{\Omega}{\sG}(t,t) \vert{u}_x\vert^2\,dx
={\frac12}  \int_{\Omega}  [{\sG}_{t}+{\sG}_{s}](t,t) \vert{u}_x\vert^2\,dx }
\\ \\ \displaystyle{\!\!\!\!\!\!\!\!
+\int_{\Omega} \, \int_0^t {\sG}_s(t,s){u}_{xt}(t)
\left[{u}_{x}(t)-{u}_{x}(t-s)\right] ds \, dx+ \int_{\Omega}
F {u}_t\, dx .}
        \end{array}\end{equation}   
        
Now, we observe that 
$$
{\frac\partial{\partial t}}[{u}_{x}(t)-{u}_{x}(t-s)]={u}_{xt}(t)-{\frac\partial{\partial s}}\left[{u}_{x}(t)-{u}_{x}(t-s)\right], \quad 0\le s\le t\le T,
$$
and then
$$
{u}_{xt}(t)
[{u}_{x}(t)-{u}_{x}(t-s)] ={\frac1 2}\left[{\frac\partial{\partial t}}\vert{u}_{x}(t)-{u}_{x}(t-s)\vert^2+
{\frac\partial{\partial s}}\vert{u}_{x}(t)-{u}_{x}(t-s)\vert^2\right]\!\!.$$
Substitution within the double integral in (\ref{form3}) gives
\begin{eqnarray*}
\label{form3b}
 &
\displaystyle{\int_{\Omega} \int_0^t {\sG}_s(t,s) \, {u}_{xt}
 [{u}_x(t)-{u}_x(t-s)] \, dx ds }\\
&
\displaystyle{
= {\frac12} \int_0^t  \int_{\Omega}
{\sG}_s(t,s) \,{\frac\partial{\partial t}}\vert{u}_x(t)-{u}_x(t-s)\vert^2  dx  ds}
\\
&
 \displaystyle{
{\frac12} \int_0^t  \int_{\Omega}
{\sG}_s(t,s) \,{\frac\partial{\partial s}}\vert{u}_x(t)-{u}_x(t-s)\vert^2  dx  ds }\\
&
\displaystyle{= {\frac12} {\frac d{dt}}\int_0^t  \int_{\Omega}
{\sG}_s(t,s)\vert{u}_x(t)-{u}_x(t-s)\vert^2  dx ds}\\
    &
    \displaystyle{ -{\frac12} 
    \int_{\Omega} \int_0^t  [{\sG}_{st}+{\sG}_{ss}](t,s)\vert{u}_x(t)-{u}_x(t-s)\vert^2   dx ds. }
\end{eqnarray*}
Taking into
account the sign conditions (\ref{G})-(\ref{G_extra}), from \eqref{form3} we obtain
\begin{equation}\begin{array}
{cl@{\hspace{0.5ex}}c@{\hspace
{1.0ex}}l}   
\displaystyle{{\frac1{2}} {\frac d{dt}}\int_{\Omega}\vert{u}_t\vert^2 dx
\,+\,{\frac1{2}} {\frac d{dt}} \int_{\Omega}{\sG}(t,t) \vert{u}_x\vert^2\,dx
\le }
\\ \\ \displaystyle{\!\!\!\!\!\!\!\!
{\frac1{2}} {\frac d{dt}}\int_0^t  \int_{\Omega}
{\sG}_s(t,s)\vert{u}_x(t)-{u}_x(t-s)\vert^2  dx ds+ \int_{\Omega}
F {u}_t\, dx .}
        \end{array}\end{equation}   
Integration  over time, in  the range $(0,t)$, $t\in(0,T)$,taking into account the sign conditions (\ref{G}) implies (\ref{ineq-lemma1}) 
and, hence, completes the proof.
\end{proof}

\section{Main results}
\setcounter{equation}{0}
According to \eqref{G}$_1$ let 
$$g_0=\min_{t\in[0,T]}\sG(t.t)>0, \qquad g_1=\max_{t\in[0,T]}\sG(t.t)>0.$$
 As a consequence of  (\ref{ineq-lemma1}) 
\begin{equation*}
\begin{split}\displaystyle 
& g_0\int_{\Omega} \vert{u}_x(t)\vert^2\, dx 
+\int_{\Omega} \vert{u}_t(t)\vert^2\, dx \le g_1\int_{\Omega} \vert{u}_{0x}\vert^2\, dx
+  \int_{\Omega} \vert{u}_1\vert^2\, dx \\
&+ 
 \int_{\Omega} \int_0^t   \vert F(\tau)\vert^2 dx \,d\tau + \int_{\Omega} \int_0^t   \vert {u}_\tau(\tau)\vert^2 dx \,d\tau ,
\end{split}\end{equation*}  
which, on application of Gronwall's Lemma, implies
\begin{equation}
\|{u}(t)\|^2_{H_0^1(\Omega)}+\|{u}_t(t)\|^2_{L^2(\Omega)}\le C e^T\,,
\label{ineq-coroll1} 
\end{equation} 
where $C=\hat C(\|{u}_0\|_{H_0^1(\Omega)},\|{u}_1\|_{L^2(\Omega)}, \|F\|_{L^2({\cal Q})})>0$.
The estimate thus obtained is needed to prove the following Theorem.

\bigskip\noindent
{\bf Theorem 4.1}. \it
Let $T>0$, $\alpha_0>0$ and \eqref{G}-\eqref{G_extra} hold. There exists a
unique solution $u$ to the problem  {\rm (\ref{eql2})-(\ref{ic+bc-G})}, subject to conditions  \eqref{data}, which satisfies 
\[  u \in C^0([0,T]; \,H^1_0(\Omega))\, \cap\, C^1([0,T];L^2(\Omega)).\]
\rm

Finally, it is easy to check that the temperature equation associated with the integral Burgers-type model also admits a unique solution.
Indeed, the application of Gronwall's Lemma as in \eqref{ineq-coroll1}, yields
\begin{equation}
\|{u}(t)\|^2_{H_0^1(\Omega)}+\|{u}_t(t)\|^2_{L^2(\Omega)}+ \int_0^t\|{u}_t(s)\|^2_{L^2(\Omega)}ds\le C e^T\,.
\label{ineq-coroll2} 
\end{equation} 
Hence we obtain the following result.

\bigskip\noindent
{\bf Remark 4.2}. \it
Under the same assumptions of Theorem 3.2, there exists a
unique solution $u$ to the problem 
\begin{equation}
\displaystyle{ \alpha_0 u_{tt}({ x},t) + \alpha_1 u_{t}({ x},t)= \sG(t,0) {u}_{xx} ({ x}, t)+ \int_{0}^{t} { \sG_{s}(t,s)~ { u}_{xx}({ x}, t-s) ~d s\, + {F}({ x},t)~ ,}}
\end{equation}
 subject to $\alpha_0,\alpha_1>0$ and conditions \eqref{ic+bc-G}-\eqref{data}, which satisfies 
\[  u \in C^0([0,T]; \,H^1_0(\Omega))\, \cap\, C^1([0,T];L^2(\Omega)).\]
\rm

\section{Conclusions}
The linear one-dimensional heat conduction problem with aging memory  (\ref{eql2})-(\ref{ic+bc-G}) is studied. It is  proved to admit a unique strong solution by means of 
suitable estimates based on the physical involved phenomena. In particular the problem is concerned about the thermal behavior of Cattaneo-Maxwell conductors as well as materials which obey the Quintanilla and Burgers-like models. 
Since the models considered here takes into account the aging of the material, this result can be applied to a wide variety of real materials. However, we limited our attention to rigid heat conductors with aging memory. This hypothesis narrows the field of applicability of the results as some aging phenomena are closely connected with the deformation history of the material. 

A possible further development consists in taking into account also the viscoelastic behavior of the body.
{The results obtained in this study represent the needed theoretical background to possible future experimental investigations. Indeed, when an existence and uniqueness result of the solution is established, then the possibility to devise and perform a numerical test is open.}
The further development we aim to consider is to generalize the results in \cite{Carillo_Giorgi} following the lines of \cite{nonrwa} to embrace the wider class of materials.

\begin{center} {\bf Acknowledgments} \end{center}
The research leading to this work has been developed under the auspices of INDAM-GNFM.  
One author (S.C.) wish to acknowledge the partial financial  support of GNFM-INDAM, INFN,  Sez. Roma1, Gr. IV - Mathematical Methods in NonLinear Physics, and SAPIENZA Universit\`a di Roma.

\end{document}